\documentstyle[aps,preprint]{revtex}

\tighten
\def\b{\begin{equation}}
\def\e{\end{equation}}
\begin{document}
\title
{On the question of trapped surfaces and black holes}
\author{Abhas Mitra}
\address{Theoretical Physics Division, Bhabha Atomic Research Centre,\\
Mumbai-400085, India\\ E-mail: amitra@apsara.barc.ernet.in}


\maketitle
\begin{abstract}
There are many observational evidences for the detection of compact
objects with masses significantly larger (in galactic cases) or much
larger (in extragalactic cases) than the upper  limits of masses of {\em
cold} Neutron Stars. Such compact objects  are commonly interpreted as
Black Holes (BHs). However, we point out that while such Black Hole
Candidates (BHCs) must be similar to BHs in many respects they, actually,
can not be BHs because existence of Black Holes would violate the basic
tenet of the General Theory of Relativity that the {\em worldline of a
material particle must be timelike at any regular region of spacetime}. We
arrive at this unique conclusion by approaching the problem from various
directions.  We feel that such ``operational'' Black Holes could be able
to explain hard X-ray tail found in the galactic BHCs because Lorentz
factor of the  plasma accreting on such objects should be considerably
higher than the corresponding NS case.
\footnote{Invited talk given in the Black Hole Workshop, Kolkata, India, 2001,
organized by Centre for Space Physics}
\end{abstract}
\section{Introduction}
This is a Workshop on ``Black Holes'' and there are many good reasons why
the present astrophysical community, in general, believes in the existence
of BHs. In several X-ray binaries there are evidences for
the existence of compact stars with masses larger than $4-5 M_\odot$, the
broad upper limit of masses of cold baryonic objects in the standard
Quantum Chromo Dynamics (QCD). Similarly there are evidences that the core
of many normal galaxies (like the Milky Way) and Active galaxies conntain
massive dark condensations with masses ranging from $10^6 - 10^{10}
M_\odot$. Currently the best explanation for such condensations is that
they are supermassive BHs. There are other physical reasons for assuming
the existence of BHs which will be mentioned towards the end of this
article.  However, we shall see that the General Theory of Relativity
(GTR) actually does not allow the existence of BHs and thus while these BHCs
must be operationally similar to BHs, in a strict sense, they must be different
from BHs. And we plead that the reader keeps an open mind about this whole
discussion.
\section {Spherically Symmetric Gravitational Field}
Any spherically symmetric metric, upon, suitable coordinate
transformation, can be brought to a form\cite{1}:
\begin{equation}
ds^2 =   g_{00} dx_0^2 +  g_{11} d x_1^2 -  R^2(x_1, x_0)(d\theta^2 +\sin^2
\theta d\phi^2)
\end{equation}
Here $x_1$ is the general radial coordinate (normally, $g_{11} <0$ and $x_0$ is the general
temporal coordinate (normally, $g_{00} >0$\cite{1}) because the chosen signature of the metric is (+1, -1, -1, -1).
Here $R$ is the {\em Invariant} Circumference Coordinate (a scalar). $R$
being a scalar, it must retain its essential property as a ``Spacelike'' coordinate.
Since, we shall deal with only radial worldlines with $d\theta = d\phi=0$,
our effective metric will be
\begin{equation}
ds^2 =   g_{00} dx_0^2 +  g_{11} d x_1^2
\end{equation}
Following Landau \& Lifshitz\cite{1}, we can rewrite it as
\begin{equation}
ds^2 =   g_{00} d{x_0}^2 [1- V^2]
\end{equation}
where
$V\equiv {\sqrt{-g_{11}} d x_1\over \sqrt{g_{00}} dx_0}$.
Here we take speed of light $c=1$. Although the physical meaning of $V$
obvious, in order to highlight that our conclusions do not depend on this
physical meaning, we will not even mention it. The element of proper time
measured by a $x_1=$ constant observer is $d\tau = \sqrt{g_{00}} dx_1$ and
the element of proper radial distance is $dl = \sqrt{-g_{11}} dx_0$.
Therefore, we may rewrite
\b
V= {dl\over d \tau}
\e
Since the worldline of a material particle
must be non-spacelike everywhere (even at a true singularity) $ds^2 \ge
0$, from Eq.(3), it follows that, we must have
$g_{00} (1-V^2) \ge 0$.
Thus if $V \le 1$, we must have $g_{00} \ge 0$, and if $V \ge 1$, we must
have $g_{00} \le 0$.
Also, the determinant of the metric coefficients $g= R^4 g_{11} g_{00} \sin^2
\theta$ is always negative\cite{1} or zero at a singularity;  or $g \le 0$, in
general.   Thus, if $V \le 1$, $g_{00} \ge 0$ and $g_{11} \le 0$. On the
other hand, if $V \ge 1$, $g_{00} \le 0$ and $g_{11} \ge 0$. Thus {\em under any circusstances}, we must have
$
-(1-V^2)/g_{11} \ge 0$.
\section{Spherical Gravitational Collapse}
If BHs  are there, they must have resulted from gravitational collapse of
stars or other fluids. The spherical collapse is studied best in the
socalled comoving coordinates, $r, t$, where $x_1=r$ is label attached to a
fluid shell with a fixed number of baryons and $x_0=t$.
It follows from the general formalism of
gravitational collapse \cite{2,3,4,5,6,7,8} that the
integration of the $0,0$ component of the Einstein equation leads to a constraint
\b
\Gamma^2 = 1 + U^2 - 2 G M(r)/R
\e
where $G$ is the gravitational constant and $M(r)$ is the gravitational
mass enclosed by a shell with $r=r$. Here the parameters
\b
\Gamma \equiv {1\over \sqrt{-g_{rr}}} {\partial R\over \partial r};\qquad~~
U \equiv {1\over \sqrt{g_{00}}} {\partial R\over \partial t}
\e
Note that while $\Gamma$ is a partial derivative w.r.t. $r$, it is a total
derivative w.r.t.  $l$ because the notion of a fixed $t$ is steeped into
the definition of proper radial length $l$. Also, while $U$ is a partial derivative w.r.t. $t$, it is a total derivative
w.r.t. $\tau$ because the notion of a fixed $r$ is steeped into the
definition of proper time\cite{2,3,4,5,6,7,8}:
\b
\Gamma= {dR\over d l};\qquad~
U = {dR\over d \tau}
\e
To fully appreciate this subtle point, the reader is specifically referred to  (see pp.
180-181 of ref. 6 and pp. 150-151 of ref. 7).
From Eqs. (5) and (8) we note that  $U$ and $\Gamma$ are interlinked as
\begin{equation}
U=\Gamma V
\end{equation}
By inserting the above relation in Eq.(6), and by transposing, we find,
\b
\Gamma^2 (1- V^2) = 1 - 2 G M(r)/R
\e
Or,
\b
{1\over -g_{rr}} \left({\partial R\over \partial
r}\right)^2 (1- V^2) = 1 - 2 G M(r)/R
\e
But in previous section we  have already found that $-(1-V^2)/g_{rr} \ge 0$.
Therefore, the L.H.S. of the Eq.(11) is $\ge 0$. So must be its
R.H.S. And hence, we must have
\begin{equation}
{2G M(r)\over R} \le 1 ; \qquad {R_{g} \over R} \le 1
\end{equation}
On the other hand, the condition for formation of a ``trapped surface'' is
that $2G M/R >1$. Thus we find that, in spherical gravitational collapse
 {\em trapped surfaces do not form}. Note that this result {\em does not depend}
on the physical interpretation of $V$ or whether $V <1$ or $V >1$, or if
there is a coordinate singularity or not.
If the collapse process  indeed continues upto $R=0$, in order that the
foregoing constraint is satisfied, we must have
$M(r) \rightarrow 0$ as $ R\rightarrow 0$.
This means that the final singularity must be of zero {\em gravitational
mass} if we assume positivity of mass.
Remember here that the quantity  $M$ is not the fixed baryonic mass :$M
\neq M_0 =m N$.
 Physically, the $M=0$ state may result when the
{\em negative gravitational energy} exactly cancels the internal energy, the
{\em baryonic mass energy} $M_0$ and any other energy.  Since trapped surface is not formed the
system keeps on radiating and $M$ keeps on decreasing till the lowest
value ($M=0$) is reached\cite{8}. We have  found that the final state corresponds
to $2GM/R \rightarrow 1$ rather than $2GM/R <1$\cite{9}. This means that the final
$M=0$ state is enclosed by a horizon and there is thus no chance of
attaining a negative mass state. We have also found that as the final
state is attained, the proper radial length\cite{9}
$
l \sim \int_0^R \sqrt {-g_{rr}} dr \rightarrow \infty
$
as $R\rightarrow 0$ and $ -g_{rr} \rightarrow \infty$. The corresponding
{\em proper time} to attain the $R=0$ state is
$
\tau \ge l/c \rightarrow \infty
$.
This means that the timelike worldlines are {\em geodesically complete}
and at any finite {\em proper time} (not just coordinate time)
sufficiently massive stars would be found as an Eternally Collapsing
Object (ECO) rather than a BH.
\section {Homogeneous Dust Collapse}
It is widely believed that by studying the problem of the collapse of the
most idealized fluid, i.e, a ``dust'' with pressure $p \equiv 0$ and no
density gradient, Oppenheimer and Snyder (OS)\cite{10} explicitly showed that
finite mass BHs can be generated. We have discussed in detail\cite{8} that
this perception is completely incorrect. For the sake of brevity, we would
like to mention here about the Eq.(36) of  OS paper which connects the
proper time $T$ of a distant observer with a parameter
$y= {R\over 2 G M}$ (at the boundary $r=r_b$) through the Eq.
\b
T \sim \ln {y^{1/2} +1\over y^{1/2} -1} + ~~ other ~ terms.
\e
In order that $T$ is definable,  the argument of this logarithmic term must
be non-negative, i.e,
$y= {R\over 2 G M} \ge 1$, or,
${2GM\over R} \le 1$,
which is nothing but our Eq.(12). Thus even for the most idealized cases,
trapped surfaces are not formed.
Since
OS assumed $M$ to be finite even at $R=0$, the final value of $y=0$ in
their case, and this may lead to non-sensical results. For example, the spatial
metric coefficient $e^\lambda$ of their paper should have been $\infty$ at the
singularity. But, it is finite in their case. Unfortunatey OS overlooked
all such inconsistencies and ended their paper soon after Eq. (36).

 However, one can still legimately wonder, if one starts with a dust of finite
 mass $M$, {\em and if the dust does not radiate}, why the condition
$2GM/R >1$ would not be satisfied at appropriate time? The point is that
if we assume $p=0$, dust is really not a fluid, it is just a collection of
incoherent finite number of $N$ particles distributed symmetrically. If
so, there are free spaces in between the dust particles and which is not the
case for a ``continuous'' fluid. Therefore although
the dust particles are symmetrically distributed around the centre of
symmetry, in a strict sense, the distribution is not really isotropic. Then the assembly of
{\em incoherent} dust particles may be considered as a collection of $N/2$
symmetric pair of particles. In GTR, a pair of particles accelerate each
other and generate gravitational radiation unmindful of the presence of
other {\em incoherent} pairs. Therefore, the gravitational mass of an
accelerating dust is really not constant! In contrast a physical spherical fluid
will behave like a coherent single body with zero quadrupole moment and
will not emit any gravitational radiation.
\section {Finite Mass Schwarzchild BH ?}
Suppose some of the readers would just refuse to accept the above conclusion
that gravitational collapse does not lead to the formation of (strict) finite
mass BH.
If so, let us assume, for the time being, the existence of a
Schwarzschild BH of mass $M$ and horigon size $R_g = 2M$ (now we set $G=1$).
The spacetime for $R \ge R_g$ is described by
\b
ds^2 = dT^2 ( 1- 2M/R) - {dR^2\over (1- 2M/R)} - R^2 (d\theta^2 + \sin^2
\theta d\phi^2)
\e
For radial geodesics of test particles, $d\theta= d\phi =0$, and it
follows that along the radial geodesic
\begin{equation}
\left({dR\over dT}\right)^2 =   {(1-2M/R)^2\over E^2} [E^2 - (1-2M/R)]
\end{equation}
where $E$ is the conserved energy per unit rest mass.
As $R\rightarrow 2M$, foregoing Eq. shows that
\b
\left({dR\over dT}\right)^2 \rightarrow (1-2M/R)^2
\e
By transposing this Eq., we find that
\b
dT^2 (1-2M/R) \rightarrow {dR^2\over 1- 2M/R} = dz^2~~ (say)
\e
By inserting the above relation in Eq.(14), we find that, for a radial
geodesic,  as $R\rightarrow 2M$
\b
ds^2 \rightarrow dz^2 - dz^2 \rightarrow 0
\e
This means that, the {\em timelike geodesic} would turn {\em null} if an
EH would exist. But this is not possible if EH is really a regular region
of spacetime. Thus, if we insist on having a BH, the only honourable
solution here would be to recognize that the EH is actually a true singularity,
i.e, the central singularity. This would mean that $R_g =0$ or $M=0$. So
if we assume that there is a BH, the permissible value of its mass is
$M=0$ or else we may have finite mass objects which could be very similar to BHs in many
ways, but which are not exactly BHs.

   Here the reader must avoid one probable confusion: Though event horizon
is a null hypersurface, here we are not concerned with the metric on this
hypersurface.  We are concerned here with the metric ($ds^2$) {\em along the
geodesic of a test particle}. The fact that that $ds^2 (EH)=0$ on the
hypersurface does not automatically mean that $ds^2=0$ along a geodesic
intersecting the EH.
\section{Kruskal Coordinates}
Since $ds^2$ is an invariant and independent of coordinate system used in
evaluating it, our above proof that if an EH would exist, the value of $ds^2$
evaluated along the geodesic of a test particle would tend to be zero,
must be valid in all coordinate systems. Yet it would be worthwhile to
obtain the same result using the Kruaskal coordinates which are believed
to be describe both the exterior and interior of a Schwarzschild BH (SBH).
 For the exterior region, we have (Sectors I and III):
\begin{equation}
u=f_1(R) \cosh
{T\over 4M}; \qquad v=f_1(R) \sinh
{T\over 4M}; \qquad f_1(R) = \pm \left({R\over 2M} -1\right)^{1/2} e^{R/4M}~~R\ge 2M
\end{equation}
And for the region interior to the horizon (Sector s II and IV), we have
\begin{equation}
u=f_2(R) \sinh
{T\over 4M}; \qquad v=f_2(R) \cosh
{T\over 4M}; \qquad f_2(R) = \pm \left(1- {R\over 2M}\right)^{1/2} e^{R/4M}; ~~R\le 2M
\end{equation}
Here the +ve sign refers to ``our universe'' and the -ve sign refers to the
``other universe'' implied by Kruskal diagram.
In either region we have
\begin{equation}
u^2-v^2= \left({R\over 2M} -1\right) e^{R/2M}~~
so~ that~~
 u^2 = v^2; \qquad R= 2M
\end{equation}
By differentiating Eq.(21) by $T$, we find
\b
2 u {du\over dT} - 2v {dv\over dT} = {e^{R/2M}\over 2M} {R\over 2M}
{dR\over dT}
\e
From Eq.(16) we see that $dR/dT =0$ at $R=2M$, hence it follows from the above
Eq.that
\b
{du\over dv} = {v\over u}; ~~R=2M
\e
Now combining Eqs.(21) and (23), we find that
\b
du^2 = dv^2; ~~ R=2M
\e
Had we differentiated Eq.(21) by $R$ or $v$ instead of $T$ we would have obtained
the same result because it can be shown that
\begin{equation}
{dv\over dR} = {rv\over 8M^2} (R/2M -1)^{-1} + {u\over 4M} {dT\over
dR}= \infty;
\qquad R= 2M
\end{equation}
In terms of $u$ and $v$, the metric for the entire spacetime is
\begin{equation}
ds^2 = {32 M^3\over  R} e^{-R/2M} (dv^2 -d u^2) - R^2 (d\theta^2
+d\phi^2 \sin^2 \theta)
\end{equation}
 Then, using Eq.(24), we  find
that for a radial geodesic
$ds^2 = 16 M^2 e^{-1} (du^2-du^2)=0$ at $ R=2M$.
 This leads us to the same conclusion: there can not be
any finite mass BH.
\section{Coordinate Singularity at EH ?}
One of the main arguments in favour of the idea  that the EH is a mere
coordinate singularity and not a physical singularity is that no
physically meaningful scalars are singular there. The most cited example
here is that of the curvature scalar
$K = {48 M^2\over R^6}$.
At $R=2M$, it is reduced to
$K= {3\over 4 M^4}$.
Thus $K$ is finite only as long as $M$ is finite. But we have found that,
for a BH, $M\equiv 0$, and hence, as is expected, actually, $K$ is
singular at the EH. In the following, let us consider another physically
meaningful scalar at the EH.
Bydifferentiating the 4-velocity, we can obtain the Acceleration 4-vector\cite{1}
\b
a^i \equiv {D u^i\over Ds} = {du^i\over ds} + \Gamma^i_{kl} u^k u^l
\e
It follows that for a radial geodesic, this acceleration vector has only
the radial component and it blows up at $R=2M$\cite{12,13}. However, this result may
be rejected by the proponents of BH hypothesis by insisting that $a^R$ is
coordinate dependent quantity and its blowing up is simply another
manifestation of ``coordinate singularity'' at the EH. But one can form a
coordinate independent scalar by contracting $a^i$\cite{12,13}:
$a \equiv  \sqrt{-a_i a^i}$.
And it is found that
\b
a = {M \over R^2 \sqrt {1 - 2 M/R}}
\e
Thus this {\em scalar indeed blows up at EH} $R=2M$. Further, $a$ would
become imaginary if there would be a spacetime beneath $R=2M$. Again all
these happenings show that the EH is the central singularity and there can
not be any spacetime beneath it. Hence, mathematically, for a
Schwarzschild BH, we must have $M=0$.
\section{Schwarzschild or Hilbert Solution}
The original paper of Schwarzschild where he worked out the spacetime around a
{\em point mass}, $M$ has recently been translated into English by Loinger
and Antoci\cite{14}. Even before this Abrams\cite{12} pointed out that, the original Schwarzschild
solution is
\b
ds^2 = dT^2 \left( 1- {2 M\over R}\right) - {dR_*^2\over \left( 1-
{2M\over R}\right)} -R^2 (d\theta^2 + \sin^2 \theta d\phi^2)
\e
where
\b
R= [R_*^3 + (2M)^3]^{1/3}
\e
The important point is that here $R$ is not the radial variable, on the
other hand the radial variable is $R_*$;
 the {\em point mass is sitting at} $R_* =0$ {\em and not at} $R=0$. Thus at,
 $R_*=2M$, obviously there is no singularity, whereas there is a central
singularity at $R_*=0$. Since there is no spacetime beneath $ R_* <0$,
there is no spacetime beneath $ R < 2M$. The prevalent Sch. solution,
which is actually the Hilbert solution is, however, still a mathematically
valid solution for a ``point mass''. So the question, which is the
physically valid solution, has to be decided on physical reasoning. Here,
some authors\cite{12,15} have rejected the Hilbert solution because because, according
to it, there would be a spacetime beneath $R <2M$ and the scalar $a$ would
then first blow up and then become imaginary. On this ground, these
authors, claim that the original Schwarzschild solution is the physically
valid solution. If so, in this picture, obviously, there would be no BH.
 On the other hand, there is a problem with the original Sch. solution; as
admitted by Sch. himself\cite{13}, in the limit of weak gravity, his equation does
not exactly reduce to a Newtonian form. On the other hand, we know that,
the prevalent Sch solution, i.e, the Hilbert solution does yield the
correct Newtonian form. Further, the angular part of either Schwarzschild or
Hilbert solution (or  any spherically symmetric metric) shows that $R$
and not $R_*$
is the Invariant Circumference Coordinate. Therefore, Hilbert solution can
not really be rejected. Then how do we resolve this paradox? We note that,
the problem lies with our premises that ``there is a {\em point mass} with
a finite gravitational mass''! And the solution lies in realizing that
 if there is a body of finite mass, it cannot be considered as a {\em
geometrical point} even at a classical level.. On the other hand, if we insist that there is a point
mass, its gravitational mass must be zero. When $M=0$, Sch. solution and
Hilbert soln. becomes identical, $R_*\equiv R$, and there is no EH, but
there is only a
central singularity in a mathematical sense. One may note here that in
classical electrodynamics there are ``point charges''. And existence of
 ``point charges'' implies singularities. On the other hand, in quantum
field theories, there are no point charges, for instance, an electron is
actually not a ``point charge''. We find here that in GTR, even at a classical, non-quantum
level, there can not be any ``point mass'' (having finite $M$).
\section{Summary and Conclusions}
We found that for
continued collapse of any perfect fluid possessing arbitrary EOS and
radiation transport properties, there is a
global constraint which demands that (exact) trapped surfaces  do not form.
 To appreciate this statement,
let us recall the following: the depth of a gravitational potential well
on a surface can be gauged by its gravitational redshift $z$. On a White
Dwarf surface, $z \sim 10^{-4}$, on a NS surface, $z \sim 10^{-1}$. But on
an  EH, $z=\infty$. And our work shows that this $z=\infty$ stage is not
realizable in any amount of finite {\em proper time} (not only in
coordinate time). So at any finite proper time gravitational collapse produces
compact objects with finite $z$ where it is possible that $z \gg 0.1$ (the
NS value). But such finite $z$ objects need not always be
static and cold, they need not represent stable solutions of equations for
hydrostatic balance (Oppenheimer- Volkoff equation\cite{16}). Luminosity of such an
object would appear reduced by a factor of $(1+z)^2$ than its lower value
(actually there would be reduction due to Doppler factor too).
And thus for a sufficiently large value of $z$, such an Eternally
Collapsing Object (ECO), may appear to approximately trap the light or radiation
emitted by it. If gravitational collapse goes on and on, since, strictly,
there is no trapped surface, the body would continue to emit radiation and
lose (gravitational) mass. So eventually, asympotically, after {\em infinite
proper time}, the body would be reduced to a $M=0$ state, and an EH would form
only at this stage.

 We found that the result that trapped surfaces do not form in GTR
collapse ($ 2GM/R \le 1$ is inscribed in the work of Oppenheimer and
Snyder too. Unfortunately they overlooked it and as a consequence their
final solution (presuming $M$ to be finite) suffers from the anomaly that
the space-space component of the metric coefficient $e^\lambda$ {\em does not blow
up even when the collapse is complete} at $R=0$ stage (this would blow up
iff $M=0$).
  In case somebody would like to ignore the above results on unspecified
reasons, we attempted to see the consequences of assuming the existence of
a finite mass Schwarzschild (rather Hilbert) BH. We considered the problem by
considering both (external) Schwarzschild coordinates and all pervading
Kruskal coordinates. In either case, we found that, if an EH would exist,
the geodesic of a test particle, which must be time like, would become,
null there. This means that the EH is no coordinate singularity but the
genuine central singularity. Technically, the mass of a SBH, thus must
be $M=0$.

 We again found that  the acceleration 4-scalar $a$
blows up at the EH indicating that EH is a true central singularity and
the mass of SBH must be $M=0$. When it is so, the curvature scalar $K=
3/4 M^4$ does blow up at the EH. Similarly, the components of the Ricci
Tensor $\sim M/R^3 \sim 1/M^2$ do blow up at the EH. And the dilemma
between the actual Schwarzschild solution and the Hilbert solution for a ``point
mass'' can be resolved only by realizing that gravitational singularities
(point masses) must have $M=0$.
 We also recall that  Rosen\cite{17}, in an
unambiguous manner noted the impossible and unphysical nature of the
$R<2M$ region:

`` so that in this region $R$ is timelike and $T$ is spacelike. However,
this is an impossible situation, for we have seen that $R$ is defined
in terms of the circumference of a circle so that $R$ is spacelike, and we
are therefore faced with a contradiction. We must conclude that the
portion of space corresponding to $R <2M$ is non-physical. This is a situation
which a coordinate transformation even one which removes a singularity can
not change. What it means is that the surface $R=2M$ represents the
boundary of physical space and should be regarded as an impenetrable
barrier for particles and light rays.''

We have tried  to show here that not only
the $R <2M$ region unphysical, it does not exist or is not ever created.
And it is well known that Einstein tried to show that BHs are not allowed
in GTR\cite{18}.
We may recall that the numerical studies of collapse
of scalar fields suggest that it is possible to have BHs of
$M=0$\cite{19}. Also, the supersymmetric string
theories find the existence of extremal BHs with charge $Q=M$,
which for the chargeless case yields $M=0$\cite{20}. Lake pointed out that
gravitational singularities could have $M=0$\cite{21}.

    Gravitational collapse of sufficiently massive bodies should indeed
result in objects which could be more compact than typical NSs ($z > \sim 0.1)$.
It is found that, if there are anisotropies, in principle there could be
static objects with arbitrary high (but finite) $z$\cite{21}. Even within the
assumption of spherical symmetry, non-standard QCD may allow existence of
cold compact objects with masses as large as $10 M_\odot$ or higher\cite{22}.
Such stars are
called Q-stars (not the usual quark stars), and they could be much more compact than
a canonical NS; for instance,
a stable  non rotating Q-star of mass 12$ M_\odot$ might have a radius of
$\sim 52$ Km. This may be compared with the value of $R_{gb} \approx
36$ Km of a supposed BH of same mass.
 And, in any
case, when we do away with the assumption of ``cold'' objects and more importantly,
staticity condition there could be objects with arbitrary high $z$.
However,  the speed of collapse of such objects,  can not be predicted by this work.
 This work can only tell
that, in principle, the collapse process can continue indefinitely because
{\em the ever incresing curvature of spacetime (Ricci Tensor) tends to stretch
the physical spacetime to infinite extent}.

In an important work,  Chakrabati and Titarchuk\cite{23} suggested that
 one of the major evidences
for the existence of BHCs is that accretion generated X-ray spectrum by
them often has a power law tail extending well above 100 KeV. Such a tail
may be understood as Compton upscattering of the soft photons around the
BHC by the infalling highly relativistic electons of the accreting plasma\cite{23}.
Recall that, although, for NS accretion, the bulk accretion speed can
reach upto $\approx 0.5 c$, the maximum bulk Lorentz factor works out to
be paltry $\gamma \sim 1.1$. Thus the flow is hardly relativistic. This
bulk comptonization requires, on the other hand, a modestly relativistic
flow with $\gamma <\sim 2$. It turns out that $\gamma = 1+z$, and an ECO
with a value of $z >\sim 1$ could be able to explain the observed hard
X-ray tail from the BHCs. On the other hand, for a EH, $\gamma =\infty$,
and since a value of $\gamma\sim few
$ is sufficient to explain the
observation, this observation may be explained by objects which are not
(strict) BHs.
   There is another line of argument for the having found the existence of
EHs in some BHCs\cite{24}. At very low accretion rates, the coupling between
electrons and ions could be very weak. In such a case most of the energy
of the flow lies with the ions, but since radiative efficiency of ions is
very poor, a spherical flow radiates insignificant fraction of accretion energy
and carries most of the energy towards the central compact object. Such a
flow is called Advection Dominated Flow (ADAF). If the central object has
a ``hard surface'', the inflow energy is eventually radiated from the hard
surface. On the other hand, if the central object is a BH, the flow energy
simply disappears inside the EH. For several supermassive BHCs and stellar
mass BHCs, this is claimed to be the case.
But   there could be several caveats in this interpretation:

 (i)  The observed
X-ray luminosities for such cases are usually insignificant compared to
the corresponding Eddington values (by a factor $10^{-5}$ to $10^{-7}$).
Such low luminosities may not be due to accretion at all. Atleast in some
cases, they may be due to Synchrotron emission.
Recently, Robertson and Leiter\cite{25} have attempted to explain the X-ray
emission from several BHCs having even much higher luminosities as
Synchrotron origin. Vadawale, Rao \& Charrabarti\cite{26} have explained one
additional component of hard X-rays from the micro-quasar GRS1915+105 as
Synchrotron radiation. The centre of our galaxy is harbours a BHC, Sgr A$^*$, of mass
$2.6 \times 10^6 M_\odot$. The recent observation of $\sim 10-20\%$ linear polarization
from this source has strongly suggested against ADAF model\cite{27}. On the other
hand, the observed radiation is much more likely due to Synchrotron process\cite{26}.
In fact, even more recently, Donato, Ghisellini \& Tagliaferri\cite{28} have shown that
the low power X-ray emission from the AGNs are due to Synchrotron process rather
by accretion process.

(ii) The x-rays if assumed to be of accretion origin, could be coming from
an accretion disk and not from a spherical flow.

(iii) Even if the X-rays are due to a spherical accretion flow, not in a
single case, we have robust independent estimate of the precise accretion rate.

(iv) Munyaneza \& Viollier\cite{29} have claimed that the accurate studies of the
motion of stars near Sgr A$^*$ are more amenable to a scenario where it is
not a BH but a self-gravitating ball of Weakly Interacting Fermions of
mass $m_f >\sim 15.9$ keV. Recall here that the Oppenheimer - Volkoff mass
limit may be expressed as
\b
M_{OV} = 0.54195 M_{pl}^3 m_f^{-2} g_f^{-1/2} = 2.7821 \times 10^9 M_\odot
(15 keV/ m_f)^2 (2/g_f)^2
\e
where $M_{Pl}= (\hbar c/G)^{1/2} $ is the PLanck mass and $g_f$ is the
degeneracy factor. With a range of  $ 13<m_f <17 $ keV, these authors
point out that the entire range of supermassive BHCs can be understood.
Note that the progenitors of the ECOs or BHCs must be much more massive (and larger in size)
than those of
the NSs.
 Then it
 follows from the magnetic flux conservation law that BHCs (at the galactic level)
 should have
magnetic fields considerably higher than NSs. It is also probable even
when they are old, their diminished magnetic fields are considerably
higher than $10^{10}$G. In such cases, BHCs will not exhibit Type I X-ray
burst activity. There may indeed be evidence for
 intrinsic (high) magnetic fields for the BHCs\cite{25}. However, in
some cases, they may well have sufficiently low magnetic field and show
Type I bursts. It is now known that Cir X-1 which was considered a BHC,
did show Type I burst, a signature of ``hard surface''. Irrespective of
interprtation of presently available observations, our work has shown that the
BHCs can not be, in a strict sense, (finite mass) BHs because then {\em timelike
geodesics would become null} on their EHs. If a BH would exist, the proper
length of an infalling astronomer or anything would become $\infty$ as it
would approach the central singularity. Then how would the observer
 stay
put in a geometrical point? Such inconsistency actually does not arise because we
have shown that one has to traverse infinite proper length over infinite
proper time in order to reach  the central singularity
($M=0$).

\vskip 0.5cm




\begin{references}
\bibitem{1} L.D. Landau and E.M. Lifshitz, {\it The Classical Theory of
Fields}, 4th ed., (Pergamon Press, Oxford, 1985).


\bibitem{2}  C.W. Misner and D.H. Sharp, {\it Phys. Rev.} {\bf 136 2B}, 571
(1964).

\bibitem{3} M. May and R. White, {Phys. Rev.} {\bf 141}, 1233 (1965).
\bibitem{4} C.W. Misner, {\it Phys. Rev.} {\bf 137 B},  1360 (1965).

\bibitem{5} W.C. Hernandez and C.W. Misner, {\it Astrophys. J.}
{\bf 143}, 452 (1966).

\bibitem{6} C.W. Misner, in {\it Astrophysics and General Relativity}, Vol.
1 (Brandeis University  Summer School Lecture), ed. M. Chretien, S. Deser,
and J. Goldstein (1968) (see pp. 180-181).

\bibitem{7} C W Misner, in {\it Relativity Theory and Astrophysics}, Vol.
3, ed. J. Ehlers (1967) (see pp. 120-121).

\bibitem{8} A. Mitra, {\it Found. Phys. Lett.}, {\bf 13}, No. 2, 543 (2000)

\bibitem{9} A. Mitra, (1998) (astro-ph/9803014).

\bibitem{10} J.R. Oppenheimer and H. Snyder, {Phys. Rev.} {\bf 56}, 455 (1939).

\bibitem{11} M. D. Kruskal, {\it Phys. Rev.}, {\bf 119}, 1743 (1960).
\bibitem{12} L.S. Abrams, {\it Can. J. Phys.} {\bf 67}, 919 (1989)
\bibitem{13} S. Antoci \& D.E. Liebscher, gr-qc/0102084
\bibitem{14} S. Antoci \& A. Loinger, physics/9905030 and 9912033
\bibitem{15} A. Loinger, astro-ph/0001453
\bibitem{16} J.R. Oppenheimer and G.M. Volkoff, {\it Phys. Rev.} {\bf 55},
375 (1939).
\bibitem{17} N. Rosen, in {\it Relativity}, (M. Carmeli, S.I. Fickler, and
L. Witten, Eds.), (Plenum, New York, 1970).

\bibitem{18} A. Einstein, {\it Ann. Math}. {\bf 40}, 922 (1939).

\bibitem{19} M.W. Choptuik, {\it Phy. Rev. Lett.}, {\bf 70}, 9 (1993).

\bibitem{20}  Gibbons G W, in {\it Gravitation and Relativity: At the turn of
the Millenium}, Ed. N. Dadhich and J. Narlikar (IUCAA, Pune, 1998).
\bibitem{21} K. Lake, {\it Phys. Rev. Lett.} {\bf 68}, 3192 (1992).
\bibitem{22} K. Dev \& M. Gleiser, astro-ph/0012265
\bibitem{23} J.C. Miller, T. Shahbaz and L.A. Nolan, {\it Mon. Not. R. Astron.
Soc.} {\bf 294}, L25 (1998).
\bibitem{24} S.K. Chakrabati \& L.G. Titarchuk, {\it Astrophys. J.}, {\bf
455}, 623 (1995)
\bibitem{25} M.R. Garcia, J.E. McClintock, R. Narayanan and P. Callanar,
{\it Astrophys. J. Lett.}, (in press), astro-ph/0012452

\bibitem{26} S.L. Robertson \& D. Leiter, {\it Astrophys. J. Lett.} (submitted, 2001), astro-ph/0102381
\bibitem{27} S.V. Vadawale, A.R. Rao \& S.K. Chakrabarti, {\it Astron.
Astrophys}, 2001 (submitted), astro-ph/0104378
\bibitem{28} E. Agol, {\it Astrophys. J. Lett.}, (in press, 2001), astro-ph/0005051
\bibitem{29} D. Dotani, G. Ghisellini \& G. Tagliaferri, {\it Astron. \&
Astrophys.} 2001 (in press), astro-ph/0105203
\bibitem{30} F. Munyaneza \& R.D. Viollier, {\it Astrophys. J.} (submitted,
2001), astro-ph/0103466

\end{references}
\end{document}